\documentclass[preprint,showpacs,preprintnumbers,amsmath,amssymb]{revtex4}

\usepackage{graphicx}% Include figure files
\usepackage{dcolumn}% Align table columns on decimal point
\usepackage{bm}% bold math

\newcommand{\be}{\begin{equation}}
\newcommand{\ee}{\end{equation}}
\newcommand{\ba}{\begin{eqnarray}}
\newcommand{\ea}{\end{eqnarray}}
\newcommand{\bt}{\begin{tabular}}
\newcommand{\et}{\end{tabular}}
\newcommand{\bc}{\begin{center}}
\newcommand{\ec}{\end{center}}
\newcommand{\ben}{\begin{enumerate}}
\newcommand{\een}{\end{enumerate}}
\newcommand{\bi}{\begin{itemize}}
\newcommand{\ei}{\end{itemize}}
\newcommand{\bmpage}{\begin{minipage}}
\newcommand{\empage}{\end{minipage}}
\newcommand{\disty}{\displaystyle}

\def \dg{\dagger}

\def\La {\Lambda}
\def\om {\omega}
\def\Om {\Omega}

\def\bul{$\bullet\;$}

\def\Hil{ {\cal H} }

\def\Hil{{\cal H} }

\begin{document}

\author{Jacek Wojtkiewicz}
\affiliation{Department for Mathematical Methods in
Physics\\Ho\.za 74, 00-682 Warsaw, Poland}
\title{
Phase diagram of the two-dimensional t-t' Falicov-Kimball model
}

\begin{abstract}
The ground-state phase diagram of two-dimensional Falicov-Kimball
model with nearest-neighbour and next-nearest-neighbour hoppings
(characterized by $t,t'$ constants, respectively) has been studied in
perturbative regime, i.e. in the case when on-site Coulomb interaction
constant $U$ is much larger than $t,t'$ ones. The fourth-order phase
diagram exhibits a rich structure, more complicated than for the
ordinary Falicov-Kimball model. Possible experimental implications of
the presence of nnn term are shortly discussed.
\end{abstract}
\pacs{71.10.Fd, 71.21.+a, 75.10.Hk, 75.30.Kz }

\maketitle

\section{Introduction}
\label{Introd}
The Falicov-Kimball model has been proposed in 1969 to description the
metal-insulator transition \cite{FK69}. Later on, it has been applied in
another important problems: mixed valence
phenomena \cite{Khom}, crystallization and alloy formation
\cite{KennLieb} and others. In this model, we are dealing with two
types of particles defined on a $d$-dimensional simple 
cubic lattice $\mathbb{Z}^d$: immobile ``ions'' and itinerant
spinless ``electrons''.
 There exist also other interpretations of the model
\cite{KennLieb}, \cite{FKRev}.

The Hamiltonian defined on a finite subset
$\La$ of $\mathbb{Z}^d$ has the form
\be
H_\La = H_{0,\La} + V_\La,
\label{hamcl0}
\ee
where
\be
\disty
H_{0,\La}=    U \sum_{x\in\La} w_{x}n_{x} - \mu_i \sum_{x\in\La} w_{x}
- \mu_e \sum_{x\in\La}n_{x},
\label{hamcl}
\ee
\be
\disty
V_\La= - \sum_{\langle x,y\rangle\in\La}t (c^\dg_{x}c_{y}
+c^\dg_{y}c_x )
\label{hamq}
\ee
Here $c^\dg_{x}$ and  $c_{x}$ are creation and annihilation
operators of an electron at lattice site $x\in\La$, satisfying
ordinary  anticommutation relations. The corresponding
number particle operator is $n_{x} = c^\dg_{x}c_{x}$.
$w_x$ is a classical variable taking values $0$ or $1$;
it measures the number of ions at lattice site $x$.
The chemical potentials of the ions and electrons are $\mu_i$ and
$\mu_e$, respectively.

The Falicov-Kimball model in its basic, ``backbone'' form given by
(\ref{hamcl}), (\ref{hamq}) is too oversimplified to give quantitative
predictions in real experiments. However, it is nontrivial lattice
model of correlated electrons and  captures  many aspects of
behaviour of such systems. It allows rigorous analysis in many
situations; for a review, see \cite{FKRev}. One can hope that a good
understanding of this simpler model might lead to better insight into
the Hubbard model, where rigorous results are rare \cite{FLU}.

One can try to make the FK model 
more  realistic by adding various
terms to the ``backbone'' hamiltonian (\ref{hamcl}), (\ref{hamq}) in the
manner analogous to that in the original Hubbard paper
\cite{Hubb}. (Other possibility is enlargement of the space of
internal degrees of freedom \cite{FKRev}, but we will not consider it
here.) The most important among them are:  consideration of
another types of lattice, particle statistics and presence of magnetic
field \cite{GruberMMU};     
correlated hopping (analysed in \cite{RefchFKM}, \cite{WL1and2},
\cite{chfk3}); 
taking into account the Coulomb interactions between nearest
neighbours, as well as (small) hopping of heavy particles
\cite{BorKot}, \cite{DFF4}; consideration of the
next-nearest-neighbour hoppings 
 (let's name this modification as the $t-t'$ model in analogy with the
 corresponding version of the Hubbard model
\cite{t-t'Hub}). This last effect has been analysed in only few
papers. In  \cite{LeJe}, authors established that if $t'\ll
t$, then the phase diagram of the $t-t'$ FKM does not differ too
much from the diagram of the pure FK model. A remarkable paper is
\cite{GruKotUel}, devoted to analysis of 
three-dimensional strongly asymmetric Hubbard model (i.e. generalized
FK one) with three hopping parameters, for large Coulomb
interaction constant $U$, in the neighbourhood of the symmetry point.
Authors have determined rigorously the 
structure of ground states and proved their thermal stability up to
terms proportional to $U^{-1} \times$ (square of the hopping
constant).  

In this paper, author examined influence of further terms
 of perturbation expansion (3-rd an 4-th ones) on the ground-state phase
 diagram  in two-dimensional situation in the  half-filling case, i.e.
 when the average value of the total particle number
 $\sum_{x\in\La}(n_x+w_x)  $ is equal to the number of sites
 $|\La|$. Effects of  
higher-order-terms turned out to be very interesting in 
 the ordinary FK model \cite{GruberMMU}, \cite{GUJ}, \cite{Ken94}.

As a first step of the study, the {\em effective Hamiltonian} has been
 derived; it can be 
written as the Hamiltonian for the Ising model with complicated
interactions, leading to strong frustration. After that, 
ground states of the Hamiltonian  have been looked for, and the phase
 diagram has been constructed. In the orders 2 and 3 it was 
possible to determine it rigorously, whereas in the fourth order use of
the {\em restricted phase diagram}  method was necessary. This method
proved their utility in the analysis of 
another versions of the FKM \cite{WL1and2},  \cite{WatsLem},
\cite{FLB}.  In this method,
one constructs collection of all periodic arrangements of ions up to
 certain values of lattice sites $N$ per elementary cell, and then one
 looks for the  configuration of minimal energy among members of this set.

As one could expect, the fourth-order phase diagram turned out to be more
complicated than in the case of the ordinary FKM.  In this last case,
one observes five phases with period 
not exceeding $N=5$ \cite{GruberMMU}, \cite{GUJ}, \cite{Ken94}; in the
case of the correlated-hopping FKM, six phases are present
\cite{WL1and2}, \cite{chfk3}. In the ground state phase diagram of
$t-t'$-FKM, {\em thirteen phases} has been found (three of them are
degenerate); the period of non-degenerate phases  does not exceed 12
sites per elementary cell. One observes also that the region occupied
by one phase (FK-like one, of density $1 \slash 4$) is {\em anomally
  large} (one can expect that this region should occupy region of the
size of the order $t^4 \slash U^3$, whereas the actual size is of the
order $t'^2 \slash U$). This phenomenon can be explained by the
lifting of the macroscopic degeneracy present in the second order by
higher-order perturbation.

Comparing phase diagrams of the FKM and $t-t'$ FKM, it turned out that
the influence of the  nnn hopping is surprisingly large.

The outline of the paper is as follows. In the Sec. \ref{sec:Hamef}, the
effective Hamiltonians up to fourth order perturbation theory  have
been derived.  In the Sec. \ref{sec:PhDgm2i3i4},
ground states and phase diagram of the effective Hamiltonians in
subsequent orders have been determined. Moreover, effects of neglected
higher-order-terms as well as temperature have been discussed.
 The last section \ref{sec:Conclus} contains summary and conclusions.

% Thermal/quantum stability, i m-potencjaly ? - NIE MA) 
% CONCLUSIONS: periodicity/non-periodicity?
%              t-t' FK jako mikroskopowy model underlying ANNNI-like?

%%%%%%%%%%%%%%%%%%%%%%%%%%%%%%%%%%%%%%%%%%%%%%%%%%%%%%
\section{Perturbation theory and effective Hamiltonian}
 \label{sec:Hamef}
%%%%%%%%%%%%%%%%%%%%%%%%%%%%%%%%%%%%%%%%%%%%%%%%%%%%%%
\subsection{Nonperturbed Hamiltonian, their ground states and phase
  diagram}
\label{subsec:Ham0}
%%%%%%%%%%%%%%%%%%%

In this paper we examine the model in the range of parameters
$t,t' \ll U$. The value of $t'$ is usually smaller than that of $t$,
however both these quantities are of the same order. 

For derivation of the effective Hamiltonian, the method worked out in
the paper \cite{DFF2} has been applied. It has this advantage that it
can serve  (provided certain  conditions
are fulfilled) as a first step to application of the 
quantum Pirogov-Sinai method and proving  thermal and quantum stability of
ground states.  Detailed description of all these procedures can be
found in \cite{DFF1} -- \cite{DFF4}. Here, only the application of the
method and results will be given, as the general scheme is identical
as in the paper \cite{DFF4}.  

To obtain the final expression, we must divide states of the system
onto {\em ground} and {\em excited} ones, and to find corresponding {\em
  projections} onto both groups. These collection of states are
identical as in  \cite{DFF4}. 

\begin{figure}
\includegraphics{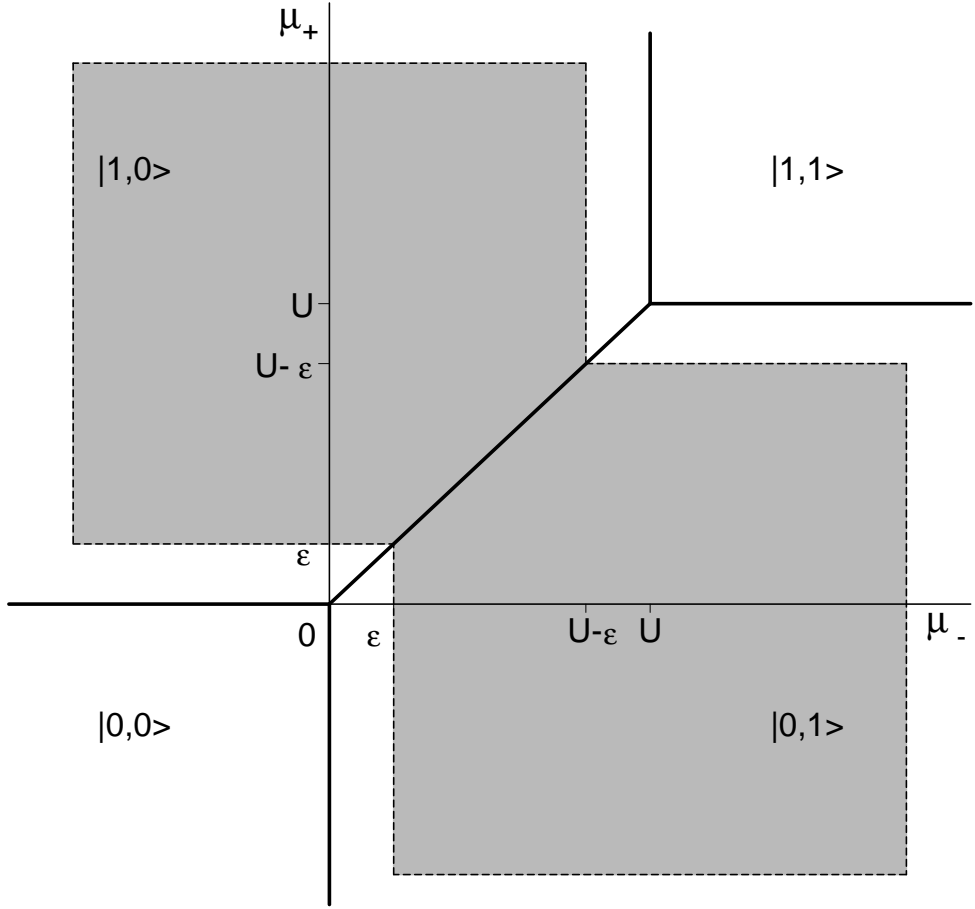}
\caption{\label{fig:fig1}  Phase diagram of the nonperturbed 
Hamiltonian (\ref{hamcl}). }
\label{fig1}
\end{figure}

Let us begin our analysis starting from the classical part of the
Hamiltonian (\ref{hamcl}); it is well known, see \cite{DFF4}.
The Hilbert space $\Hil_i$  on the $i$-th site is 
 spanned by the  states: $|n_{i,+},n_{i,-}\rangle$ or,
explicitely, $|0,0\rangle$,
$|1,0\rangle$, $|0,1\rangle$ and  $|1,1\rangle$.
The corresponding energies are: $0; - \mu_+; - \mu_-; U-\mu_+ - \mu_-$.
 The phase diagram consist of the following  four regions. 
In region $I$, defined by $\mu_+<0,\mu_-<0$, all sites are empty. In
two twin
regions $II_+, II_-$ given by conditions:
$II_+$:
 $\mu_{+}>0$, $\mu_{+} > \mu_{-}$, $\mu_{-}<U$ (for $II_-$, one should
interchange the subscripts $+$ and $-$)
all  sites are in the  $|1,0\rangle$
(corresp. $|0,1\rangle$) state. In the region $III$, given by:
$\mu_+>U, \mu_->U$, all sites are doubly occupied.  This situation is
illustrated on Fig. \ref{fig:fig1}.

%Fig. 1

We choose the states $|1,0\rangle$ and $|0,1\rangle$ as ground
 states. They are divided from excited ones by energy gap $\Delta= {\rm
 min}(U, \mu_+, \mu_-)$. It means that we analyse the phase diagram 
in some subset of the region  $II_+ \cup II_-$ (the shaded region on
 the Fig. 1). The most interesting situation takes place in the
 neighbourhood 
 of the  $\mu_+ = \mu_-$ line between regions $II_+$ and $II_-$; on
 this line, we observe a macroscopic degeneracy. 

 The projection operator on ground states at $i$-th site is
\be
P^0_i=(n_{i,+}-n_{i,-})^2
\label{ProjFK}
\ee

%%%%%%%%%%%%%%%%%%%%%
\subsection{Effective Hamiltonians up to  4-th order of perturbation
  theory}
\label{subsec:Hamef4}
%%%%%%%%%%%%%%%%%%%%%

Expression for effective Hamiltonian in fourth-order perturbation
theory for the ordinary FK model can be found in 
\cite{DFF4}, Table 2. The 4-th order
effective Hamiltonian for $t-t'$ FKM can be derived using the same
methodology, described in \cite{DFF4}, Sec.~3. (It should be stressed
that  expressions up to 4-th order have been derived, for the
3d model, in the paper \cite{GruKotUel}. Unfortunately,  authors
didn't analyse effects of orders 3 and 4). 

 As a final result of calculations, one
obtains, after specialization to the half-filled case (i.e. the
situation where the total number of particles is equal to the number
of lattice sites):

\bul Second-order correction: 
\be
 H^{(2)}_{0} = -h \sum_{i}S_i +
\sum_{d(i,j)=1}\frac{t^2 }{2U}(4S^3_i S^3_j - {\bf 1})
+ \sum_{d(i,j)=\sqrt{2}}\frac{t'^2 }{2U}(4S^3_i S^3_j - {\bf 1})
\label{H2_0}
\ee
where:  $h =\mu_i - \mu_e$; $S_i$ is the classical one-half spin on
the  lattice site $i$;
it is related to the variable $w_i$ by the formula:
 $S_i=(w_i-1)\slash{ 2}$.

\bul Third-order correction:
\be
 H^{(3)}_{0} = \frac{t^2 t'}{U^2} \sum_{i,j,k}
\left( 6 \, S_i S_j S_k -\frac{1}{2} ( S_i +  S_j + S_k) \right)
\label{H3_0}
\ee
where summation is performed over all triples $\{ i,j,k\}$ on the
lattice such that $\{i,j\}$ and $\{j,k\}$ are nearest neighbour bonds 
forming the angle $\pi\slash 2$. 

\bul Fourth-order correction is the most complicated one and is a sum
  of two-body (2b) and four-body (4b) interactions:
\[
 H^{(4)}_{0} =C +
J^{\rm 2b}_1 \!\! \sum_{d(i,j)=1} \!\! S^3_i S^3_j
+J^{\rm 2b}_2  \!\!\sum_{d(i,j)=\sqrt{2}} \!\! S^3_i S^3_j
+J^{\rm 2b}_3  \!\!\sum_{d(i,j)=2} \!\! S^3_i S^3_j +
J^{\rm 2b}_4 \!\! \sum_{d(i,j)=\sqrt{5}}  \!\!S^3_i S^3_j +
J^{\rm 2b}_5 \!\! \sum_{d(i,j)=\sqrt{8}} \!\! S^3_i S^3_j
\]
\be
+J^{\rm 4b}_1 \!\!\sum_{\pi_1(ijkl)} \!\! S_i S_j S_k S_l
+J^{\rm 4b}_2 \!\!\sum_{\pi_2(ijkl)} \!\! S_i S_j S_k S_l
+J^{\rm 4b}_3 \!\!\sum_{\pi_3(ijkl)} \!\! S_i S_j S_k S_l
+J^{\rm 4b}_4 \!\!\sum_{\pi_4(ijkl)} \!\! S_i S_j S_k S_l
\label{H4_0}
\ee
In formulas above, we have: 
$C=3\tau\slash 2 + 5 \tau' + 3\tau''\slash 2$;
 $J^{\rm 2b}_1=-18\tau - 32\tau'$;
 $J^{\rm 2b}_2= 6\tau -36\tau' -18 \tau''$;
 $J^{\rm 2b}_3=4\tau - 4\tau' + 6\tau''$;
 $J^{\rm 2b}_4=12\tau'$;
$J^{\rm 2b}_5= 4\tau''$;
 $J^{\rm 4b}_1 = 40\tau + 80\tau'$;
$J^{\rm 4b}_2 = 40\tau'' $;
 $J^{\rm 4b}_3 = 40\tau' $;
 $J^{\rm 4b}_4 = 40\tau' $, where: $\tau=t^4\slash U^3$;
 $\tau'=t^2 t'^2\slash U^3$; $\tau''=t'^4\slash U^3$. Sets
 $\pi_\alpha(ijkl)$  are defined in the following way:  $\pi_1$
 (``square'') is formed by spins 
 occupying  vertices  $(0,0)$, $(0,1)$, $(1,1)$ and $(1,0)$;  $\pi_2$
 (``diagonal  square'') is formed
 by:  $(0,0)$, $(1,1)$, $(0,2)$ and $(-1,1)$; $\pi_3$ (``big
 triangle''):  $(0,0)$, $(0,1)$, $(0,2)$ and $(1,1)$; $\pi_4$:
 (``rhomb'')  $(0,0)$, $(1,0)$, $(2,1)$ and $(1,1)$. The summation
 over four-body interactions in 
 (\ref{H4_0}) is performed over all sets obtained from plaquettes
 $\pi_1, \dots, \pi_4$ above
 by operations compatible with lattice symmetries (translations,
 rotations by multiple of $\pi/4$, reflections, inversions); the
 plaquette $\pi_\alpha(ijkl)$ occupies sites $i,j,k$ and $l$.
    
%%%%%%%%%%%%%%%%%%%%%%%%%%%%%%%%%%%%%%%%%%%%%%%%%%%%%%%%%%%%%%%%%
\section{Ground state phase diagrams in order 2 and 3}
\label{sec:PhDgm2i3i4}
%%%%%%%%%%%%%%%%%%%%%%%%%%%%%%%%%%%%%%%%%%%%%%%%%%%%%%%%%%%%%%%%%
\subsection{Order 2}
\label{subsec:rzad2}
%%%%%%%%%%%%%%%%%%%%%
The ground-state phase diagram of the system described by the Hamiltonian
(\ref{H2_0}) can be obtained by rewriting the Hamiltonian in the
following equivalent form:
\be
 H_0^{(2)} = \sum_{\pi_1(ijkl)} h^{(2)}_{0;ijkl}
\label{H2_mp_gen}
\ee 
where
\be
 h^{(2)}_{0;ijkl} = \frac{t^2}{U}(S_i S_j + S_j S_k + S_k S_l
 + S_l S_i) + \frac{2t'^{\,2}}{U}(S_i S_k + S_j S_l)
-\frac{h}{4} (S_i  + S_j + S_k+ S_l)  
\label{H2_mp}
\ee
(lattice sites $i,j,k,l$ are arranged anticlockwise on the plaquette). 

\begin{figure}
\includegraphics{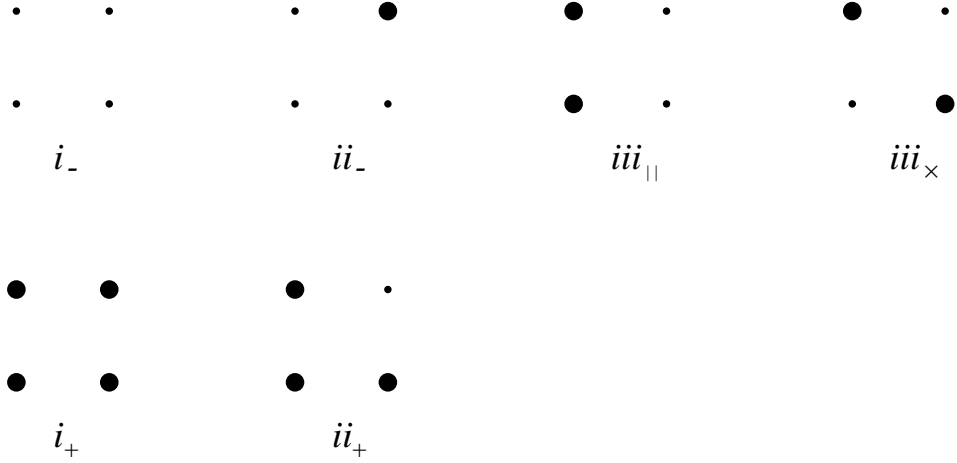}
\caption{\label{fig:fig2}  Possible configurations of $2\times 2$
  plaquettes. For plaquettes $ii$ and $iii$ one should take into
  account plaquettes obtained from those illustrated above by
  rotations.  Small dots denote empty lattice sites (or spins
  ``down'' in the spin language), big dots -- occupied lattice sites
  (spins ``up'', respectively).}
\label{fig2}
\end{figure}

\begin{figure}
\includegraphics{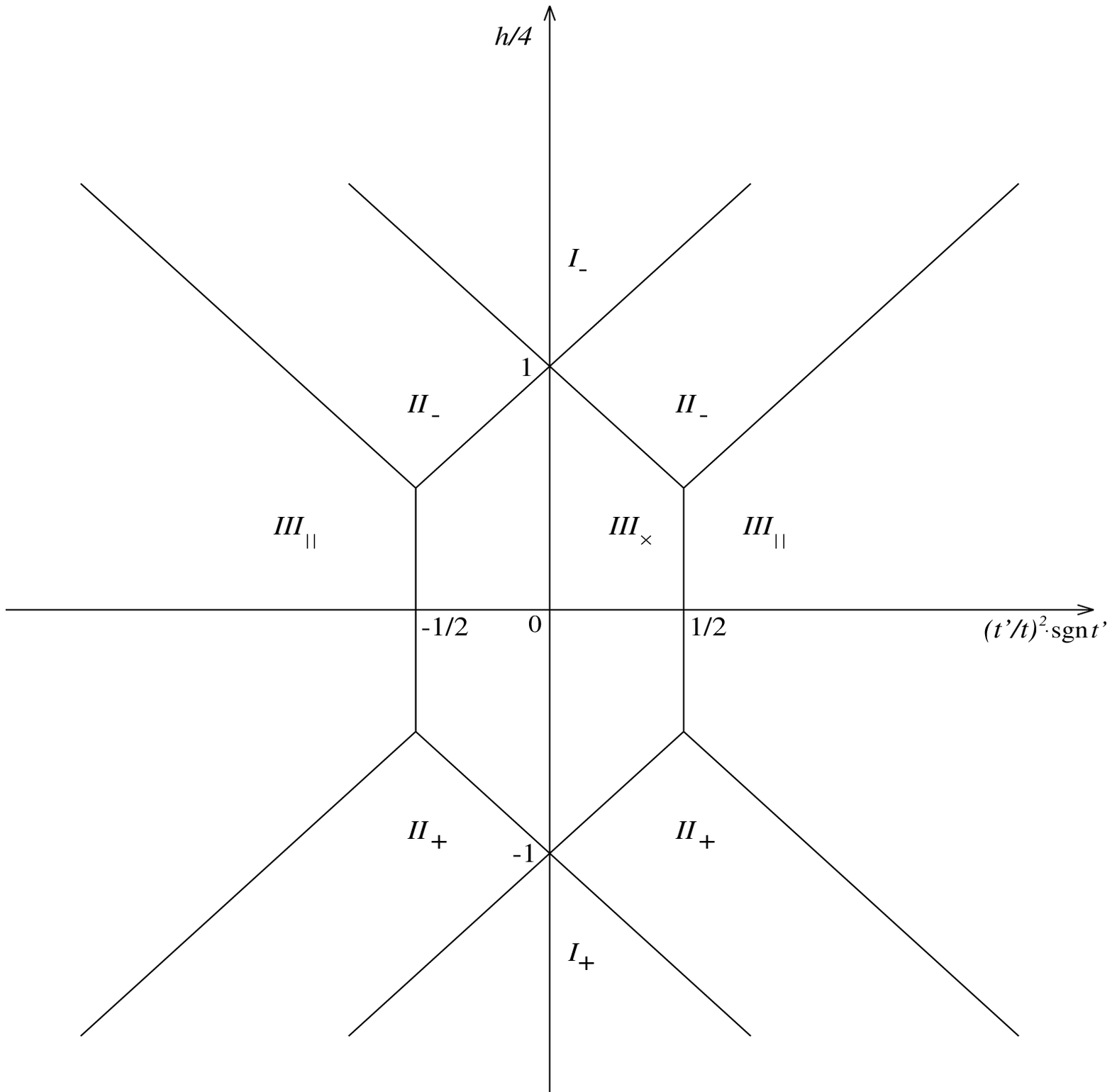}
\caption{\label{fig:fig3}  Ground state phase diagram in the second
  order perturbation theory for the Hamiltonian (\ref{H2_mp_gen}),
  (\ref{H2_mp}).  $h$ is the difference of chemical
  potentials. Phases $I$ and $III$ are unique (modulo rotations and
  translations), whereas phases $II$ are macroscopically
  degenerate. In order 3, the phase diagram is a small deformation of
  the above picture.} 
\label{fig3}
\end{figure}

It is easy to check that the Hamiltonian rewritten in the form
(\ref{H2_mp}) is an {\em m-potential} (\cite{DFF1},  \cite{DFF2},
\cite{Slawny87}; this definition is also reminded in the
Appendix). In such a case, we can replace the process of minimization
of energy over the whole lattice by the problem much simpler: the
minimization of energy over the set of plaquette configurations. 
These configurations are presented on Fig.~\ref{fig:fig2}.  It leads
to the picture  of the phase diagram as illustrated on
Fig.~\ref{fig:fig3}. This 
diagram possess  two obvious symmetries. One of them
 is due to symmetry of the hamiltonian  (\ref{H2_0}) with respect
  to the change of sign $t' \to -t'$; the phase diagram is also
  symmetric with respect to such a change of sign. The second one is the
  symmetry of phase diagram with respect to the change $h \to -h$;
  however, in this case, one should also replace  configurations by their
  mirror images (i. e. $S_i \to - S_i$).
%Fig. 2

%Fig. 3

 Phases $I_+$ (full) 
and $I_-$ (empty; for illustration, see configuration 0 on the
Fig.~\ref{fig:fig4}) are build from plaquettes  $i_+$ and $i_-$,
respectively. These phases are {\em unique}. We have similar situation
for  regions $III_\times$ (N\'{e}el phase; Fig.~\ref{fig:fig4},
configuration 10) and $III_{||}$ ( Fig.~\ref{fig:fig4}, configuration
11 (it is an analogon of the ``planar'' phase in
\cite{GruKotUel}). They are build from plaquettes  $iii_\times$ and $iii_{||}$,
respectively. Again, these phases are unique (modulo
translations). The situation for phases $II_+$, $II_-$ differs from
previous ones. These phases are build from plaquettes $ii_+$, $ii_-$,
respectively (see  Fig.~\ref{fig:fig4}, configuration No. 5 as an
example); however, they are {\em non-unique} and  possess 
macroscopic degeneration. One can easily check that there is a large
dose of freedom in building of  lattice configurations from these
plaquettes. We encounter here  situation similar  to this which
happens for the antiferromagnetic Ising model on triangular lattice.  

%%%%%%%%%%%%%%%%%%%%%%%%%%%%%%
\subsection{Order 3}
\label{subsec:rzad3}
%%%%%%%%%%%%%%%%%%%%%%%%%%%%%%

Now, let us check how the phase diagram will change under switching
third-order terms on. Let us rewrite the third-order correction
(\ref{H3_0}) in the equivalent ``plaquette'' form:
\be
 H_0^{(3)} = \sum_{\pi_1(ijkl)} h^{(3)}_{0;ijkl}
\ee 
where
\be
 h^{(3)}_{0;ijkl} = \frac{6\,t^2t'}{U^2}(S_i S_j S_k + S_j S_k S_l + S_k
 S_l S_i + S_l S_i S_j) 
-\frac{3\,t^2 t'}{2U^2} (S_i  + S_j + S_k+ S_l)  
\label{H3_mp}
\ee
(again spins $i,j,k,l$ are arranged anticlockwise on the plaquette). 

The full Hamiltonian, up to third-order terms, is the sum of terms
(\ref{H2_mp}) and (\ref{H3_mp}). As in previous Subsection, one can
check that it is an m-potential. Moreover, it turns out that the
presence of third-order terms {\em does not modify ground states of
  plaquettes}. In the other words, plaquette configurations which were
ground-states in the second order, remain ground states also in third
order! This implies that degeneracy of phases
 $II_+$ i $II_-$ {\em is not lifted} and they still are
 degenerate. What {\em does} change, it is location of the boundary
 between phases. The difference in location of phase boundaries  in
 orders 2  and 3 is of the order $t^2t'\slash U^2$. 

The phase diagram in third order possess certain kind of
symmetry.  It is discussed in more details in the next Subsection. At
this moment, we only
 conclude  that {\em the phase diagram in 3. order is a
small deformation of the second-order phase diagram}. 

%%%%%%%%%%%%%%%%%%%%%%%%
\subsection{Phase diagram in fourth order}
\label{subsec:rzad4}
%%%%%%%%%%%%%%%%%%%%%%%%
Regions occupied by phases $II_\pm$ in both second and third order
exhibit macroscopic degeneracy. One can expect that they will be
sensitive against perturbations and that in some of next orders this
degeneracy will be lifted. This happens yet in fourth order; we
describe the situation below.

This picture has been obtained by the restricted phase diagram
method. Recall that  in this method, 
one takes into account all periodic configurations up to certain values
of lattice sites $N$ per elementary cell, and then one minimizes the
energy over this set of configurations. 

\begin{figure}
\includegraphics{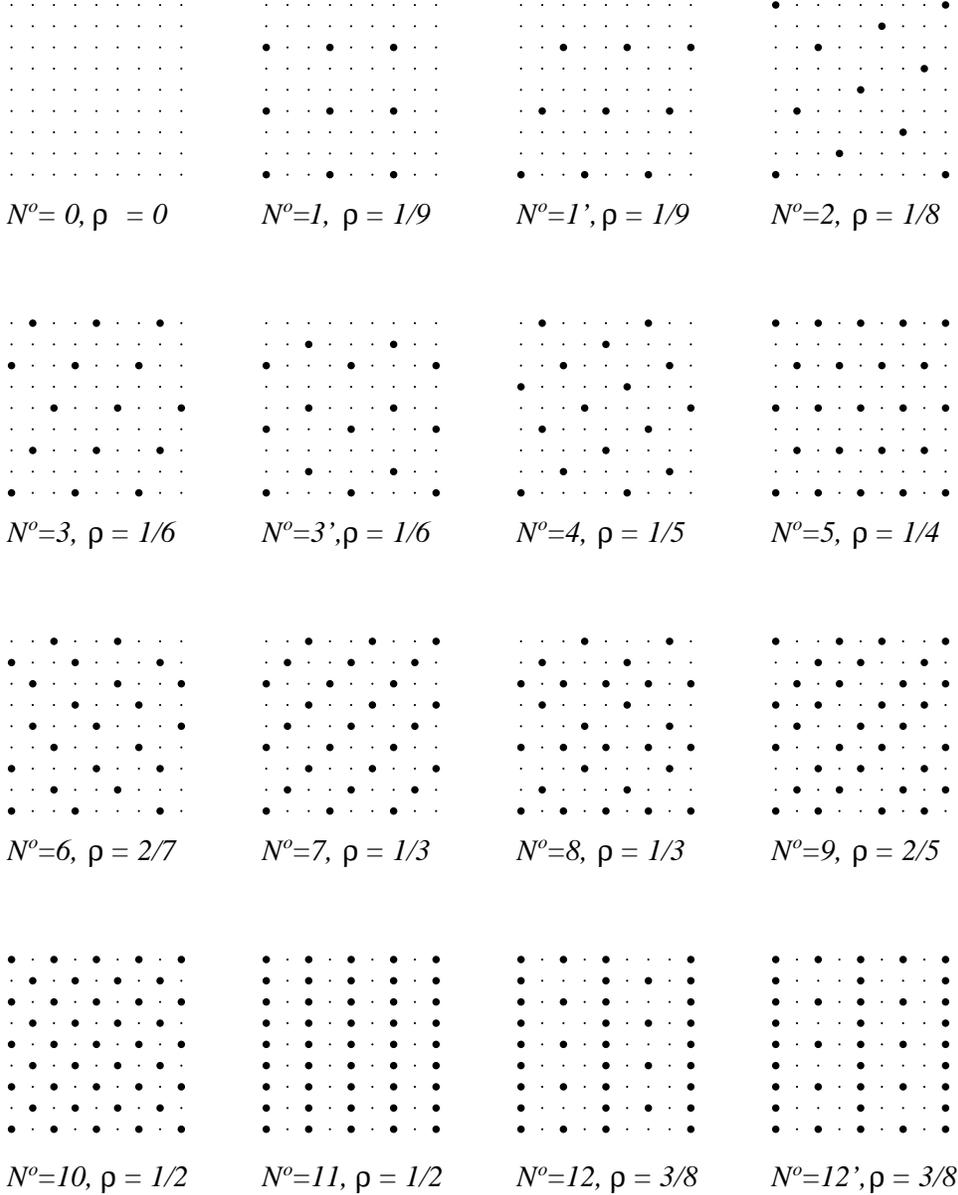}
\caption{\label{fig:fig4}  Phases appearing on the ground-state phase
  diagram for the fourth-order effective Hamiltonian (\ref{H4_0}).
  Configurations: 0,2,4-11 are unique modulo lattice
  symmetry operations. Configurations possessing densities $1\slash
  9$, $1\slash 6$ and $3 \slash 8$ are degenerate and form (infinite?)
  series; two first examples for every such a series are shown (1,1'
  etc.). } 
\label{fig4}
\end{figure}

\begin{figure}
\includegraphics{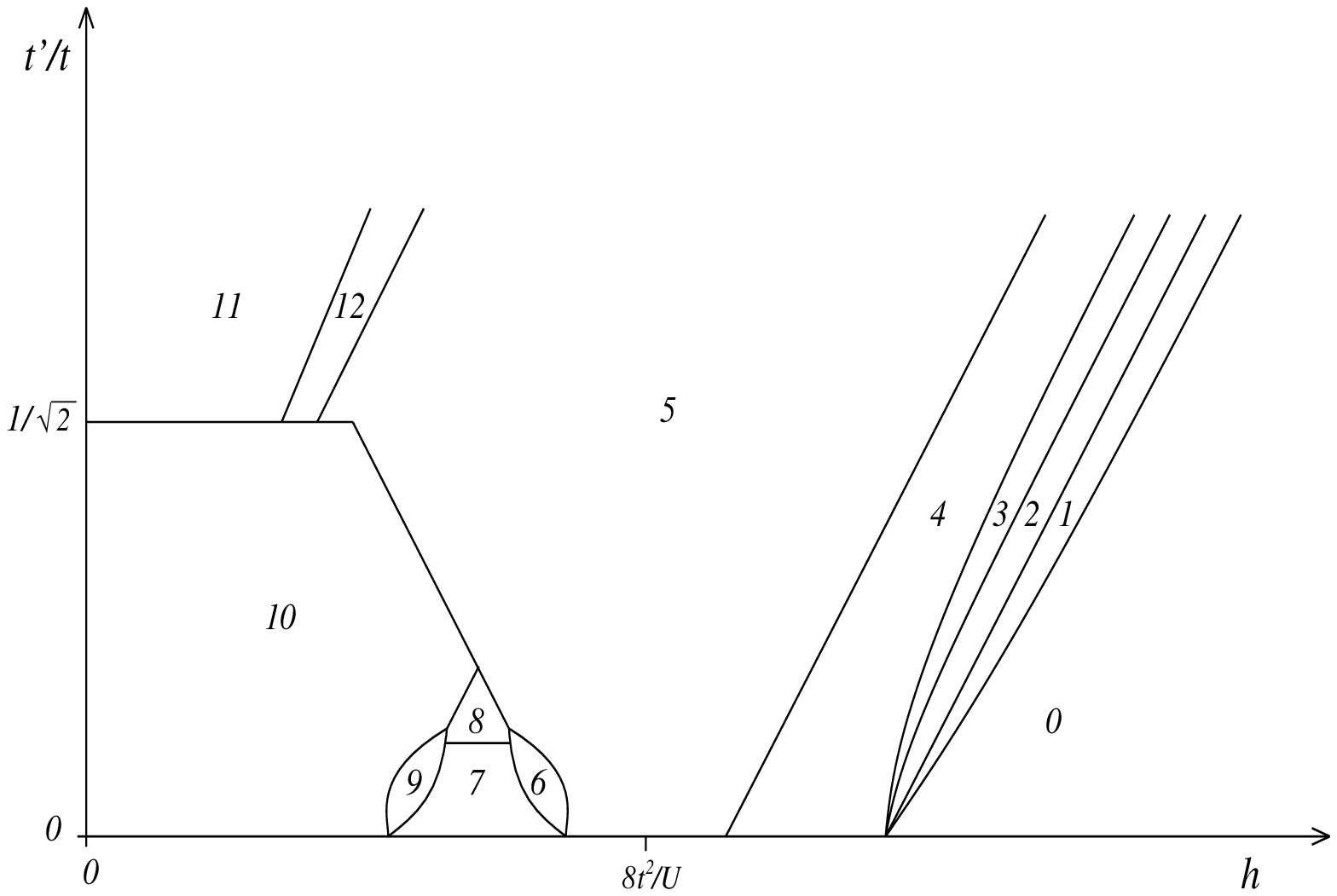}
\caption{\label{fig:fig5} The ground-state phase
  diagram for the fourth-order effective Hamiltonian
  (\ref{H4_0}). Only topological aspects of the phase diagram are
  displayed, as most of phases occupy very narrow regions. Phases:
  1-4, 6-9 and 12: width of these regions is of fourth order in
  expansion parameter; other phases occupy regions of width of the
  second order in expansion parameter. Only one
  quarter  (i.e. values $h>0, t'>0$)  are shown,
  as for other quarters, topological structure of the phase diagram is
  the same.} 
\label{fig5}
\end{figure}

The ground-state phase diagram is much more complicated than for the
ordinary FKM  -- see Figs.~\ref{fig4} and~\ref{fig5} but it is still
manageable  (in some respects it  is similar to the
phase diagram of the FKM on triangular lattice, studied in
\cite{GruberMMU}).  Thirteen phases have been detected on phase
diagram; it seems that three of them are  degenerate. Moreover, for each
phase present for $h>0$, there corresponds their ``mirror'' for $h<0$
(this mirror is obtained by the
change of occupied sites onto unoccupied ones and vice versa). Periods
of these phases do not exceed 12 sites per elementary cell. Such a
picture emerged at $ N=12 $ and hasn't changed up to 
$N=27$ (which corresponds to more than $3\times 10^7$ trial
configurations). 
For this reason, author claims that this phase diagram is
``exact but non-rigorous''. 

%Fig. 4

%Fig. 5

The phase diagram in fourth order possess certain
(pseudo)symmetries. Let us describe the situation order-by-order.

In the second order, we have the Hamiltonian symmetric with respect to
the change $h\to -h$ and $t'\to -t'$. It is obvious that the phase
diagram is symmetric with respect to the change $h\to -h$, if one
change also configuration to its mirror image.

In the third order, the Hamiltonian is no longer symmetric. However,
their ground states are easily determined, because the 
Hamiltonian is 
expressible as a sum of m-potentials, defined on $2\times 2$
plaquettes. It turns out that on the phase diagram, there are present
{\em the same} phases as in the second order. The difference between
phase diagrams in the second and third order is apparent in location
of  phase boundaries; corresponding lines are shifted by a factor
proportional to $tt'^2\slash U^2$. This shift is {\em symmetric} with
respect   to the change $h\to -h$. In the other words: If some
boundary  between phases $i$ and $j$ is shifted by
$\epsilon$ for $h>0$, then, for $h<0$, the boundary between mirrors of
phases $i$ and $j$ is shifted  by $-\epsilon$. 

Let us analyse (pseudo)symmetries of the fourth order phase
diagram. Consider first some third-order configuration of minimal
energy for $h>0$ (call it CME${}^3_+$); for every such a
configuration, we have its mirror, which is also the configuration of
minimal energy for some  $h<0$ (call it CME${}^3_-$). Now, let us  add the
fourth order contribution to the Hamiltonian. This term is symmetric with
respect to reversing of spins;
so, if for $h>0$ we have some configuration of minimal energy
CME${}^4_+$, then for $h<0$ the configuration of minimal
energy  CME${}^4_-$ will be the mirror of CME${}^4_+$. 

Let us summarize the above by the statement that the topological
structure of phase diagram is the same for $h>0$ and $h<0$; for every
phase $i$ appearing for $h>0$, we have corresponding mirror $\hat{i}$
for $h<0$. Phase boundaries between phases $i$ and $j$ and their
mirrors  $\hat{i}$ and $\hat{j}$ are related by:
\[
h_{i\slash j} = A t^2 + A' t'^2 + B t^2 t' + C t^4 + C' t^2 t'^2 + C'' t'^4
\]
\[
h_{\hat{i}\slash \hat{j}} = -A t^2 - A' t'^2 + B t^2 t' - C t^4 - C'
t^2 t'^2 - C'' t'^4 
\]
Let us stress that this situation (i.e. the same structure of phase
diagram for $h>0$ and $h>0$ is a very peculiar property of the
fourth-order Hamiltonian (\ref{H2_0}), (\ref{H3_0}), (\ref{H4_0});  it
is due to degeneracy and (pseudo)symmetries of lower-order
Hamiltonians. For more general Hamiltonian, we have no such similarity.
 
Perhaps, the most interesting effect of the presence of the term with
next-nearest-neighbour hopping is the appearance of the anomally large
region occupied by the phase 5 (FK-like phase with density $1\slash
4$). At first sight, phases appearing in the fourth-order perturbation
theory should occupy region of the width $p(t,t')\slash U^3$ (where
$p(t,t')$ is some homogeneous fourth-order polynom in
$t$ and  $t'$). However, it turns out that phase 5 occupies a region of
width proportional to $t'^2\slash U$, i.e. of the same order as the
N\'{e}el phase, appearing in the second order! One can explain this
phenomenon in the following way:  
Regions occupied by phase  $II_+$ (the situation with phase $II_-$ is
analogous)    has width of the order $t'^2\slash U$
proportional to exhibit macroscopic degeneracy
both in 2-nd and in 3-rd order. The fourth-order
perturbation lifts this degeneracy and  as a result  the phase
$II_+$ (an 'ancestor')  transforms into non-degenerate phase 5 (the
'descendant')  of the same density, and occupies region approximately
as large as an 'ancestor' $II_+$. 

Most of phases is unique, but there are also phases which remain
degenerate even in the 4-th order (phases 1, 3, 12). Strictly
speaking, the restricted phase diagram method detects here only  {\em
  finite} degeneracy, i.e. finitely many ground states with
identical energy and density but different orderings (we don't count
trivial 
degeneracy due to symmetry operations, i.e. translations, rotations
and reflections). Number of these ground states grows with $N$; for
$N=27$ we observed:  eight  phases of equal energy and density $1\slash 9$,
(the first two such configurations are phases  1 and 1'); eight phases
of density  $1 \slash 6$ (the first two such configurations are phases  
3 and 3'); five phases of density  $3 \slash 8$ (the first two of them are
12 and 12'). Moreover, it has been observed that every member of such
collection of phases with equal density is build up from {\em
  identical} plaquette configurations $3\times 3$. They can be 'glued'
together in various arrangements and there is no uniqueness in such a
procedure, i.e. resulting lattice configuration is
non-unique. Situation here is similar to that which happens in
the second order for phases  $II_\pm$. It is natural to conjecture
that we encounter here the {\em macroscopic degeneracy}, i.e. presence of
{\em infinite} number of configurations of identical density and
energy. However author (so far) can't prove this.

Phase diagrams in orders 2 and 3 are rigorous (by writing out
Hamiltonians as sums of m-potentials). Author tried to do analogous
thing in the fourth-order by an attempt to construct m-potentials in a
manner analogous as in \cite{Ken94},
\cite{GruberMMU}, \cite{chfk3}, however, it succeeded only for some
phases, but not for a whole phase diagram. 

An analysis above concerned the 'truncated' phase diagram, i.e. the
phase diagram of the fourth-order effective Hamiltonian
(\ref{H4_0}). Which changes can result as an effect coming from next
(neglected) orders and temperature? Author expects (and conjectures)
that  most phases (i.e. all with exception of degenerate ones:
i.e. 1, 3, 11 and their descendants) are {\em stable} ones
\cite{DFF1}, \cite{DFF2}. For such phases,  {\em Peierls conditions}
(both classical and quantum ones -- see \cite{DFF1}, \cite{DFF2})
are fulfilled, and regions occupied by these phases deform in only
small manner upon   thermal and
quantum perturbations.  This assertion concerns regions of phase diagram
sufficiently far from phase boundaries. For regions of width of the
order of $t^5\slash U^4$ around phase boundaries, one cannot formulate
any statements without going into next orders of perturbation
theory. However, at present author cannot prove stability and Peierls
conditions. (It could be proved by construction of m-potentials, which
failed so far).  Regions
occupied by degenerate phases constitute different problem. Here,
changes caused by thermal and quantum perturbations can be very
significant; they are ``terra incognita'' and we skip this subject,
leaving it  as an open  problem.   

%%%%%%%%%%%%%%%%%%%%%%%%%%%%%%%%%%%%%%%%%%%%%%%%%%%%%%%%%%%
\section{Summary and conclusions}
\label{sec:Conclus}
%%%%%%%%%%%%%%%%%%%%%%%%%%%%%%%%%%%%%%%%%%%%%%%%%%%%%%%%%%%

The effective Hamiltonian and phase diagram for ground states of the
$t-t'$ FKM have been 
determined up to fourth order of perturbation theory. In the second
and third order, phase diagram was constructed by rewritting the
Hamiltonian as a sum of m-potentials. The phase
diagram in the fourth order has been determined by the method of
restricted phase diagrams; author claims that such a picture is
``exact but non-rigorous''. The phase diagram is considerably more
complicated than for the ordinary FKM, but still it is
manageable. Thirteen phases are present (plus their ``mirrors'');
three of these phases are degenerate (i.e. possess identical density
and energy but different ordering). Author conjectures that this
degeneracy is macroscopic (the method of restricted phase diagrams
handles only finite number of configurations).

Let us list some of the features of the phase diagram in fourth order:
\begin{enumerate}
\item For small $t'$, the phase diagram is similar to this for the
   ordinary    FKM; however, for larger $t'$, these phase diagrams are
   quite   different. 
\item One observes anomally large region occupied by one of the phases
  appearing in 4-th order (phase number 5). It could have experimental
  implications as a possibility of appearance of ``charge density
  waves'' more exotic than N\'{e}el ordering. 
\item The phase diagram of the full model (i.e. with inclusion of
  neglected terms perturbation theory, as well as temperature) is more
  difficult to examine than for the ordinary FKM. At this stage of
  investigation, almost nothing rigorous can be said. However, author
  expects that after switching the quantum and thermal perturbations
  on, the phase diagram will change in only small manner inside
  regions occupied by nondegenerate phases (0,2,4--10,12). But regions
  of width of the order $t^4\slash U^3$ around the phase boundaries,
  as well as regions occupied by degenerate phases 1,3,11, are out of
  possibilities of present analysis.  
\item The Falicov-Kimball model is sensitive to perturbations. 
  Modifications of the Hamiltonian such as introduction of correlated
  hopping, nnn-hopping or consideration of the FKM in non-perturbative
  regime  (i.e. for values
  of $t\slash U$ not very small) can significantly or even drastically
  modify the phase   diagram.  
\end{enumerate}
 
%%%%%%%%%%%%%%%%%%%%%%%%%%%%%%%%%%%%%%%%%%%%% 
\appendix
%%%%%%%%%%%%%%%%%%%%%%%%%%%%%%%%%%%%%%%%%%%%%
\section{ Ground states, m-potentials} 
%%%%%%%%%%%%%%%%%%%%%%%%%%%%%%%%%%%%%%%%%%%%%

 {\em The Hamiltonian} $H_\La$ , where $\La\subset \mathbb{Z}^d$, is a
function defined on $\Om_{|\La|}$ -- the space of all configurations
of the system. Usually Hamiltonian is defined as a sum
of  {\em potentials}, i.e. functions defined on subsets of  $\La$:
$H_\La=\sum_{B\subset \La}\Phi_B$. Usually one imposes restrictions
such that potentials are {\em finite-range} ones, i.e. such that
$|B|\leq M$,  $M$ finite. It is also assumed that potentials are
translation invariant. 

 {\bf m-potential.} Now, consider the system on an infinite
lattice ${\mathbb{Z} }^d$. Assume that sets  $B_a$ (i.e. potential
supports) are translation of a fixed plaquette $B$ by a lattice vector
$a$: $B_a=\tau_a B$, where $\tau_a$ is an operator of such
translation.   We say that the
function $\Phi_B$ is an  {\em m-potential}, if there exist
configuration  (perhaps, non-unique) $\om^0\in\Om_N$ with the
following properties: {\em i)} For every plaquette $B_a$, the ``plaquette
energies'', i.e. values of the Hamiltonian calculated on the plaquette
 $B_a$: $\Phi_{B_a}(\om^0)$ are all equal; {\em ii)} For every another
configuration  $\om\in\Om_N$, the condition: $\Phi_{B_a}(\om^0)
\leq \Phi_{B_a}(\om)$ is fulfilled.  

 {\bf Ground states of the classical Hamiltonian.} If there
exist such configuration  $\om^0\in\Om_N$ as above, then we call it
{\em the ground state} of the Hamiltonian. 

The property of the potential to be an m-potential can be reformulated
as follows. If a given potential is an m-potential, then the {\em
local} minimality of energy (i.e. minimality on a plaquette) implies
the {\em global} minimality on the whole lattice. The property that a
potential is an m-potential is  very important one, as it
replaces searching of ground states of the infinite lattice by looking
for the minima on a finite set. Unfortunately, some given potential
possess this property only exceptionally. (But fortunately, in
Secs. \ref{subsec:rzad2} and \ref{subsec:rzad3} they {\em share}  
such property!)
 A method to avoid this
obstacle is to find -- for a given potential $\Phi$ -- an {\em equivalent}
potential $\Phi'$, such that $\Phi'$ is an m-potential. However, in
general it is difficult task.
%%%%%%%%%%%%%%%%%%%%%%%%%%%%%%%%%%%%%%%%%%%%%%%%%%%%%%%%%%%%%%%%%

\noindent {\bf Acknowledgements.}

\noindent The author is grateful for numerous discussions with
 R. Lema\'{n}ski and  J. Mi\c{e}kisz. This work was
supported by  the 
Polish Research Committee (KBN) under  Grants: No.~2~P03B~131~19 and
110\slash 501\slash SPUB\slash 27, and
by the Postdoctoral Training Program HPRN-CT-2002-00277.
%%%%%%%%%%%%%%%%%%%%%%%%%%%%%%%%%%%%%%%%%%%%%%%%%%%%%%%%%%%%%%%%%

\end{document}